# GdN Nanoisland-Based GaN Tunnel Junctions

*Sriram Krishnamoorthy[†], Thomas F. Kent[‡], Jing Yang[‡], Pil Sung Park[†], Roberto C. Myers[‡,†], Siddharth Rajan[†,‡]\**

[†]Department of Electrical and Computer Engineering and [‡]Department of Materials Science and Engineering, The Ohio State University, Columbus, OH 43210, United States

ABSTRACT

**Tunnel junctions could have a great impact on Gallium Nitride and Aluminium Nitride based devices such as light emitting diodes and lasers by overcoming critical challenges related to hole injection and p-contacts. This paper demonstrates the use of GdN nanoislands to enhance inter-band tunneling and hole injection into GaN p-n junctions by several orders of magnitude, resulting in low tunnel junction specific resistivity (1.3 X $10^{-3}$ $\Omega$-$cm^2$) compared to the previous results in wide band gap semiconductors. Tunnel injection of holes was confirmed by low temperature operation of GaN p-n junction with a tunneling contact layer, and strong electroluminescence down to 20K. The low tunnel junction resistance combined with low optical absorption loss in GdN is very promising for incorporation in GaN-based light emitters**

The low ionization efficiency of p-type dopants and low hole mobility lead to high parasitic resistances in III-nitride optoelectronics that cause significant losses in devices such as energy-efficient solid state lighting LEDs, laser diodes. Interband tunnel junctions could improve the efficiency and functionality of a variety of III-nitride semiconductor devices for visible and ultra violet emitter and photovoltaic applications.[1-7] N-type tunneling contacts to resistive p-type layers could lead to significant improvement in the efficiency of light emitting diodes (LED), lasers, and solar cells, while device structures such as multi- junction solar cells and multiple active region LEDs also require tunnel junctions as a critical component. The resulting efficiency improvements in LEDs for energy efficient solid state lighting could have global energy impact.

However, in wide band gap materials such as GaN, it is extremely challenging to achieve interband tunneling between p-type and n-type regions. The larger band gaps in this system and dopant solubility limitations lead to thicker depletion regions, large energy barriers to tunneling and higher tunneling resistance than in lower gap semiconductors, and traditional heavily doped p-n junctions[8] can therefore not provide low tunneling resistance. Recently, the spontaneous and piezoelectric polarization[9,10] in the III-nitride system using structures like GaN/AlN/GaN, [3-5] GaN/InGaN/GaN,[11-13] and AlGaN/GaN[14] were used to overcome limitations imposed by the large energy band gap. While GaN/InGaN/GaN tunnel junctions with very low resistance ($10^{-4}$ $\Omega$-cm$^2$) have been demonstrated,[6] the dependence on crystal polarity and the high absorption coefficient of InGaN limit the application of this approach in devices.

Mid-gap states created by intentional defects can be used to overcome tunneling resistance limitations imposed by constraints such as the energy gap and dopant solubility by providing intermediate states that can reduce the tunneling width between p and n regions in a tunnel diode, as shown in Figure 1a. In the GaAs system, this idea was successfully used to improve tunneling using low temperature (LT) molecular beam epitaxy (MBE) grown GaAs[15], where native point defects provided the mid-gap states, and epitaxial semi metallic ErAs[16-18] and MnAs[19] nanoparticles in GaAs, where the nanoscale islands provide intermediate states to assist in tunneling. In this paper, we show for the first time that rare earth nitride (GdN) nanoislands can create mid-gap states in a p-n junction that enhance interband tunneling by *several orders of magnitude*.

Bulk GdN[20,21] is a cubic rock salt semiconductor with a theoretical indirect band gap of 0.7-0.85 eV[21] and (111)-oriented GdN has a 9.4% lattice mismatch with the basal plane of wurtzite GaN. GdN nanoislands have successfully been demonstrated on c-plane GaN,[22] additionally, it was shown that high quality single crystalline GaN layers could be overgrown on these nanoislands[22] with no new observable extended defects in the overlayer. The formation of GdN as nanoislands is important because, GaN overgrown on planar (111) GdN[23] was found to be polycrystalline due to broken rotational symmetry along the c-axis. The ability to grow high quality crystalline GaN on top of GdN enables us to embed nanoislands within a matrix of single crystal GaN, while maintaining high crystalline quality in the GaN. Furthermore, the indirect gap of GdN leads to negligible absorption at visible wavelengths (see supporting information), which is a critical criterion for incorporation into optoelectronic devices. GdN nanoislands therefore provide a promising way to create mid-gap state-enhanced tunnel junctions in the III-nitride system.

GaN $p^+$-GaN/GdN nanoisland/$n^+$-GaN junctions as depicted in Figure 1b were grown along N-face (000-1) and Ga-face (0001) orientations using $N_2$ plasma-assisted molecular beam epitaxy (PAMBE). GdN coverage was incomplete and deposition of 2.4 ML of GdN resulted in a partial coverage with 3 nm tall islands oriented along the [111] direction on wurtzite [0001] GaN.[22] Details about island formation dynamics and transmission electron microscope images of the islands are described elsewhere.[22] A 50 nm p-type GaN was grown on top of the GdN nanoislands, and was found to be single crystalline based on in-situ reflection high-energy electron diffraction, as expected from previous work.[22] The structure was capped with a 30 nm heavily p-type doped GaN layer to enable formation of low resistance ohmic contacts. The device was processed using contact lithography, with Ti (20nm)/Au (200nm) ohmic contact for the GaN:Si layer and Ni(20nm)/Au(200nm) ohmic contact on $p^{++}$ GaN cap. Additional details of the growth conditions used in this work is presented in the supporting information.

The principle of GaN/GdN/GaN tunnel junction operation is explained here. Under forward bias, the holes in the valence band of p-GaN and electrons in the conduction band of n-GaN tunnel into states in GdN and recombine. Since the forward bias current does not rely on direct band to band tunneling(from n-GaN to p-GaN), we do not expect band-crossing related negative differential resistance (as expected in a normal heterojunction Esaki diodes). In reverse bias, electrons in the valence band of p-GaN tunnel into conduction band of n-GaN via states in GdN.

The temperature dependent electrical characteristics of the Ga-polar GaN/GdN/GaN TJ (20 μm X 20 μm) are shown in Figure 1c. At a reverse bias of 1 V, the current density changes from 7.8 A/cm$^2$ at 77 K to 78 A/cm$^2$ at 300K. The IV curves in the low temperature, high current regime had a quadratic dependence, indicating that space charge limited transport may be playing a role.[24] This shows that after hole freeze out (which occurs at relatively high temperatures due to the high ionization energy of Mg), holes are injected through the tunnel junction and travel across the p-doped region. Further details are in the Supporting information.

The transport mechanism in p-GaN/GdN/n-GaN TJ device includes tunneling into states in GdN and subsequent recombination. Since GdN is an indirect band gap material, phonons are expected to play an important role in the tunneling processes, which would result in temperature dependence in the tunneling current. The Fermi level in the p- and n- regions would depend on the temperature, and thereby modify the barrier heights for tunneling. In addition, the freeze-out of holes in the p-type region can significantly change the series resistance, thereby increasing resistance with lowering the temperature. With regards to the recombination processes, temperature could play an important role in band to band recombination. However, if trap-assisted or surface assisted recombination dominates (as it would in a nanostructure), recombination processes could be significantly enhanced. The exact transport mechanism involved in GaN/GdN/GaN is currently under investigation and would necessitate investigations on the exact band line ups at GdN/n-GaN and GdN/p-GaN heterojunctions, which are not available at present.

Figure 2a shows the room temperature electrical characteristics of the Ga-polar and N-polar TJ devices. At a reverse bias of 1 V, the Ga-polar GaN/GdN/GaN device had a current density 78 A/cm$^2$, while the same epitaxial layer grown along the N-polar orientation had a comparable current density of 99 A/cm$^2$. This indicates that GaN/GdN/GaN tunnel junctions have relatively weak dependence on crystal orientation and polarization in the reverse direction. However, in the forward bias, as shown in Figure 2a, the N-polar TJ shows higher current density than the Ga-polar TJ. This is because the polarization field in GdN is in the same direction as the depletion field in case of N-polar TJ. This effect is in fact quite similar to reduced turn–on voltage reported in previous work on N-polar green LEDs.[25]

The relatively weak dependence of the tunneling current on crystal orientation and polarization in the reverse direction can be understood from the band diagram of the structures. To calculate the energy band diagram of a p-GaN/3 nm GdN/n-GaN heterojunction, we model GdN as a non-polar material with a narrow bandgap[21], and assume that the polarization discontinuity at the GaN/GdN interface leads to an interface charge density that is equal to the spontaneous polarization sheet charge of GaN. This is depicted in the charge diagram in Figure 2b, for the Ga-face and N-face oriented p-n junctions. For p-GaN up structure, the field due to the spontaneous polarization sheet charge dipole $\sigma_{sp,GaN}$ is in the same direction as that of the depletion field in case of N- polar, where as the direction is opposite in the case of Ga – polar structure. The effect of doping in GaN on the band diagram is simulated using a self consistent Schrödinger Poisson solver[26], and is shown in Figure 2c,d, for Ga-polar and N-polar case respectively, assuming equal conduction band offset and valence band offset and a dielectric permittivity[21] of 7. The depletion width is expected to be lower in the N-polar case than in the Ga-polar case, with the difference being less significant as the doping density is increased. At the highest doping density shown ($N_D = N_A = 5 \times 10^{19}$ cm$^{-3}$), the depletion field exceeds the polarization field and the band diagram looks similar for both the orientations, and the depletion width is not significantly different in the two cases.

To confirm tunneling injection of holes, we designed an n-GaN/GdN/p-GaN/n- GaN structure (sample A) as shown in Figure 3a. This structure uses an n-GaN/GdN/p-GaN tunnel junction to contact the p-GaN layer of a GaN p-n junction, thereby eliminating the need for a metal p-contact. As the p-n junction is forward biased, the GaN/GdN tunnel junction is reverse biased. Electrons from the valence band of p- GaN tunnel into n-GaN, leaving a hole behind in the p-GaN layer. This is equivalent to tunnel injection of holes from an n-GaN contact into the p-type material, and so if the tunnel junction is ideal, we expect this structure to behave like a simple p-n junction. The two important requirements for such a tunnel junction to be incorporated in a LED or solar cell would be that (1) the voltage drop across the tunnel junction during the device operation is negligible, with appreciable tunneling close to zero bias, and (2) the specific resistivity of the tunnel junction is minimal. The structure with the GaN/GdN TJ contact layer (sample A) was compared with a structure with a reference $p^+$-$n^+$ GaN TJ contact layer (sample B) that has the same structure as sample A, but without the GdN nano island layer.

Atomic force microscope image (Figure 3b) of the sample surface shows typical step flow growth morphology with a low rms roughness of 0.6 nm. The electrical characteristics of sample A shown in Figure 3c show a GaN p-n junction behavior with rectification, indicating that the tunnel junction behaves as an ohmic contact to the p-GaN layer. The series resistance of the device with GaN/GdN TJ (Sample A) in forward bias was calculated from a fit of the linear portion of the forward bias characteristics to be $1.3 \times 10^{-3}$ $\Omega\text{-cm}^2$. This includes the contact resistance, series resistance in the p-GaN and n-GaN regions, and the tunnel junction resistance. The top contact resistance was measured using transfer length method (TLM) method, and found to be $1.3 \times 10^{-5}$ $\Omega\text{-cm}^2$. The estimated series resistance of the thin p-GaN ($\sim 10^{-5}$ -$10^{-6}$ $\Omega\text{-cm}^2$) and n-GaN layers ($\sim 10^{-7}$ $\Omega\text{-cm2}$) used in this study are negligible compared to the measured total forward bias resistance. Hence, the specific tunnel junction resistivity can be extracted to be $1.3 \times 10^{-3}$ $\Omega\text{-cm}^2$. The forward turn on voltage of this device with the incorporation of a tunnel junction is the sum of the voltage drop across the forward biased p n junction and the voltage drop across the reverse biased tunnel junction. At a reference current density of 20 A/cm², the p contact-less p-n junction with GaN/GdN TJ (Sample A) had a voltage drop of 3.05 V. At a typical current density of 20 A/cm$^2$ the voltage drop across the GaN/GdN TJ would be 26 mV (calculated from the extracted resistivity value), which is negligible compared to the turn-on voltage of the GaN p-n junction. For comparison, I-V characteristics of the p$^+$-n$^+$ GaN TJ device (Sample B) are also shown in Figure 3c, and a voltage drop of 4.5 V measured at a current density of 20 A/cm$^2$. This indicates an additional voltage drop of 1.45 V across the p$^+$-n$^+$ GaN TJ, which is similar to past reports where GaN-based tunnel junction LEDs led to an extra voltage drop ranging from 0.6 V to 1.5 V[1,2]. These results demonstrate that GdN nanoislands can provide very efficient hole tunneling into p-GaN layers.

The temperature dependent electrical characteristics of sample A are shown in Figure 3d. Even at 77 K, where freeze out of holes is expected, high current density was observed in forward bias confirming efficient injection of non-equilibrium holes into the layers at low temperature. We also observed blue-UV electroluminescence at both low temperature and room temperature, as shown in Figure 3e, confirming injection of holes. At 20K, the band to band (3.4eV), and band to acceptor (3.2eV) emission in GaN was observed with two additional peaks marked as peak A and peak B. The origin of these peaks could be related to defect levels in GaN, or emission from GdN, or absorption of UV photons in GdN and reemission. However, the strong electroluminescence signal obtained at low temperature, indicates efficient hole injection due to the tunnel junction even in the absence of any thermally ionized holes.

To benchmark these results against the literature, the specific tunnel junction resistivity achieved in different material systems are compared in Figure 3f.[3,5,27-29] It is evident from the figure that tunnel resistivity increases exponentially with band gap, and that while low resistivity in wider band gap materials such as GaN is an inherent challenge, heterostructure engineering using GdN nanoparticles provides a route to overcome this limitation. The low tunnel resistivity (1.3 X $10^{-3}$ $\Omega$-$cm^2$) achieved in this work would lead to negligible tunnel junction voltage drop loss for the typical drive current in LEDs and solar cells. Further investigations are necessary to understand the true band line-up, interfacial dipoles, and related effects in these heterostructures.

Optimization of the island size, density, and growth conditions could help to reduce the resistance further, which would enable incorporation in high brightness LEDs. For applications in lasers which require higher current density (~ kA/$cm^2$), further optimization of these tunnel junctions will be needed. P-contacts with resistance of $10^{-4}$ $cm^{-2}$ can be achieved after significant process optimization in university and industry labs. Tunnel junctions may approach or exceed the efficiency of such p-contacts with optimization. However, the availability of a tunnel junction enables epitaxy above the contact, enabling very important device designs that include transparent n-type spreading layer and multiple active region devices[6]. With the availability of tunnel junctions for both Ga-polar and N-polar oriented layers, several new directions for GaN electronics and optoelectronics are possible. Several groups have indicated the advantages of N-polar P-up[25,30,31] or Ga-polar P-down[32,33] LEDs. However, P-down LEDs have been limited due to the large p-type layer spreading resistance, and therefore need a tunnel junction to connect the p-layer to an n-type substrate. Heterostructure polarization engineering based approaches using InGaN are not effective since they work for N-polar oriented P-down p-n tunnel junctions[6] and not for a Ga-polar oriented P-down p-n junctions. The alternate approach (using GaN/AlN/GaN) could be used for the aforementioned cases, but does not provide low enough tunneling resistance to be useful. The GdN approach presented here overcomes these limitations by enabling both Ga-polar and N-polar oriented p-n junctions. GdN nanoislands could also be embedded in wide band gap barriers, combining polarization and nanoparticle engineering for lower resistance.

Previous work on GaN TJ LEDs[1,2,34] showed improved current spreading leading to enhanced light output. However the TJ LED devices had an additional voltage drop across the TJ. The efficient low resistance tunnel junction achieved in this work can be incorporated to enhance the performance of top emitting LEDs. The incorporation of low resistance tunnel junctions could enable III-nitride vertical cavity emitters[35] where current spreading and resistance of p-type layers are major issues, while magnetic properties of GdN[20] enable spintronic devices that benefit from the high spin lifetimes in GaN.[36] The design of GaN bipolar devices such as p-n diodes and BJTs has been limited by p-type contacts and resistance. Low resistance tunnel junctions can now be used to re-design these bipolar devices to achieve greater performance or functionality. Finally, since GdN-based junctions do not require polarization, they could also be used to improve performance for emerging non-polar and semi-polar GaN devices.

In summary, we demonstrated that epitaxial GdN nanoislands can be used for efficient inter-band tunneling in GaN p-n junctions on Ga- and N-polar orientations. Efficient tunnel injection of holes from n-type GaN top contact into p-type material was achieved, and the tunnel junction specific contact resistivity was estimated to be $1.3 \times 10^{-3}$ $\Omega$-cm$^2$. This demonstration of efficient tunneling using GdN nanoislands, together with previously reported polarization-engineered tunnel junctions[3-5] could lead to designs for III-nitride electronic and optoelectronic devices with higher performance and functionality.

TOC Graphic

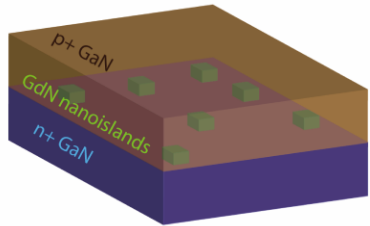 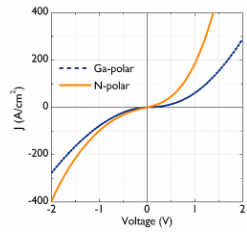

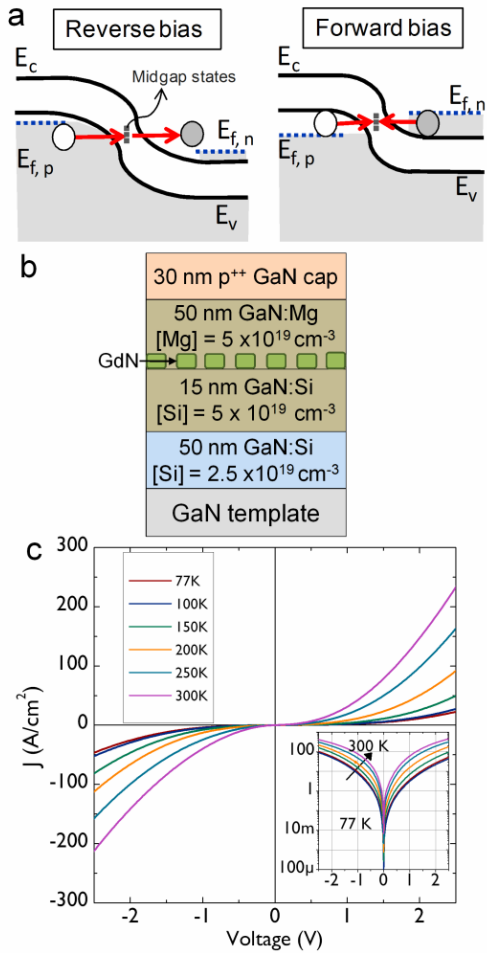

**Figure 1.** (a) Schematic showing operation of a mid gap states assisted tunnel junction in forward bias (electron hole recombination at mid gap states) and reverse bias( two step tunneling process through the mid gap states). (b) Epitaxial stack of GaN/GdN/GaN inter-band tunnel junction. The concentration mentioned in the figure refers to the doping density. (c) Temperature dependent electrical characteristics of Ga-polar GaN/GdN/GaN TJ Inset: Temperature dependent Log I- V characteristics

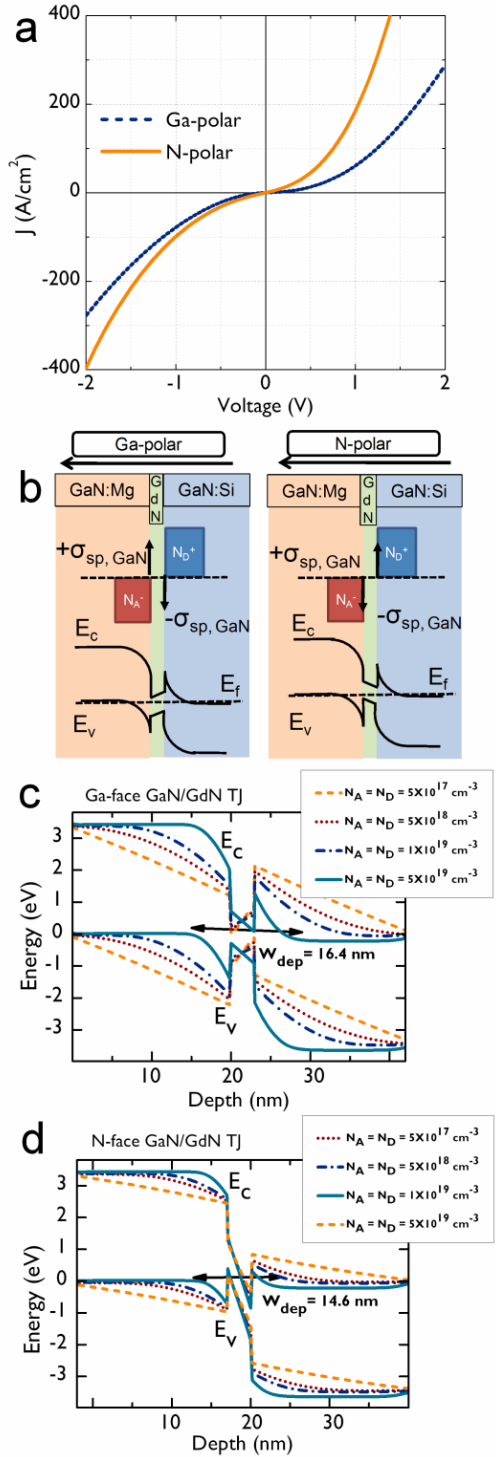

**Figure 2.** (a) I-V characteristics of Ga-polar and N-polar GaN/GdN/GaN TJ. Reverse characteristics of the Ga-polar and N-polar TJ shows similar current density indicating a weak polarization dependence on tunneling resistance. (b) Equilibrium charge profile and band diagram of Ga-polar, N-polar oriented p-GaN/ GdN/ n-GaN TJ. Polarization and depletion fields

are aligned for the N-polar orientation case. (c) Effect of increased doping in the case of Ga-polar and (d) N-polar p- GaN/ GdN/ n- GaN TJ.

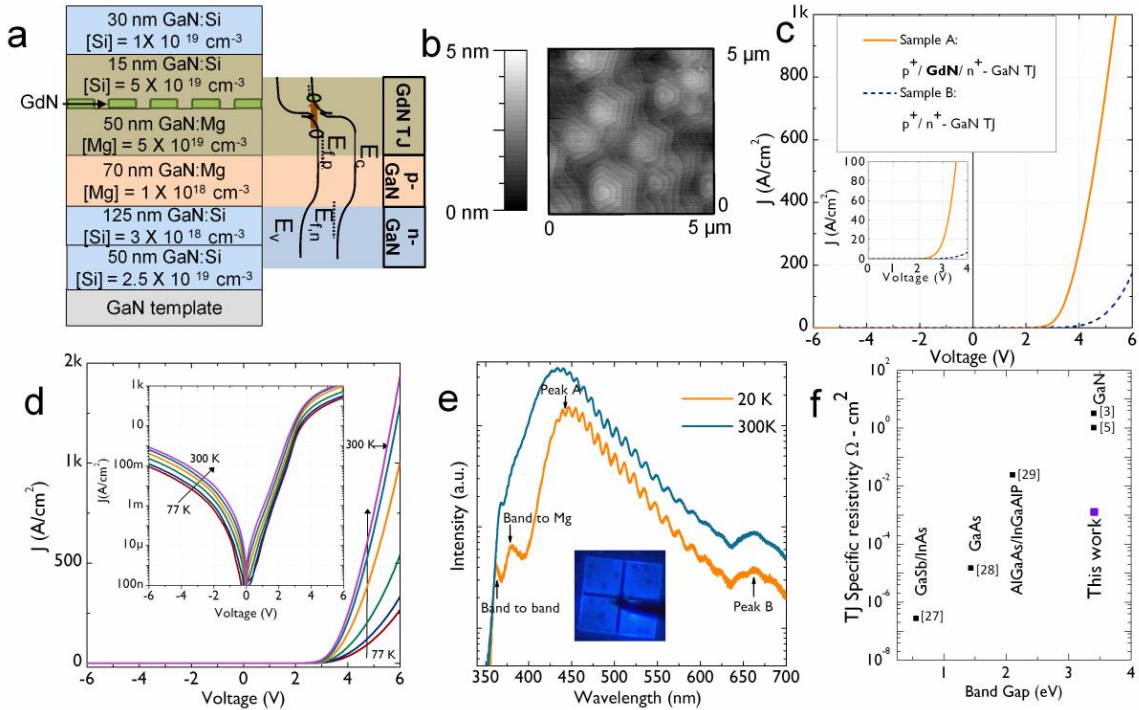

**Figure 3.** (a) Epitaxial stack of the p-contactless p-n junction where a GaN/GdN TJ connects the p-region with the n-region. The energy band diagram shows tunnel injection of holes into p-type material from the n-type contact when the p-n junction is forward biased. (b) Atomic force microscope image showing smooth surface morphology (rms roughness = 0.6 nm) (c) I- V characteristics of the p-contactless p-n junction with and without GaN/GdN TJ. The sample with GaN/GdN TJ (Sample A) has negligible voltage drop across the tunnel junction, while the reference sample (Sample B) has a significant resistive drop. Inset:I-V characteristics showing the low forward current density regime. (d) Temperature dependent electrical characteristics of Sample A  Inset: Temperature dependent Log J- V characteristics of sample A, indicating high forward bias current even at low temperature. (e) Electroluminescence spectra obtained from sample A at 20K and room temperature. The low temperature EL shows GaN band to band as well as band to acceptor emission peaks. The additional peaks observed are marked as peak A and peak B. Inset: Optical micrograph picture of the device showing uniform blue emission.(f) Plot of specific tunnel junction resistivity achieved in different material systems as a function of band gap. The work presented here represents very low tunneling resistance achieved in GaN

ASSOCIATED CONTENT

**Supporting Information**. Discussion of materials and methods, space charge limited transport in GaN/GdN/GaN tunnel junctions, and absorption due to GdN nanoislands. This material is available free of charge via the Internet at http://pubs.acs.org.

AUTHOR INFORMATION

**Corresponding Author**

\* Email: rajan@ece.osu.edu

ACKNOWLEDGMENT

The authors would like to acknowledge Anup Sasikumar, Prof. Emre Gur and Prof. Steven A Ringel for low temperature measurements shown in this work. We acknowledge funding from ONR DATE MURI program (Program manager: Dr. Paul Maki).